\documentclass{article}
\usepackage{graphicx}
\usepackage{amsfonts,amsmath,cite}
\usepackage[ansinew]{inputenc}
\usepackage{a4wide, url}

\newcommand{\hpaa}{\hat{P}^{a}_{\alpha}}
\newcommand{\hpaae}{\hat{P}^{a}_{\alpha_1}}
\newcommand{\hpaat}{\hat{P}^{a}_{\alpha_2}}
\newcommand{\hpaas}{\hat{P}^{a}_{\alpha_3}}
\newcommand{\hpbb}{\hat{P}^{b}_\beta}
\newcommand{\hpbbl}{\hat{P}^{b}_{\beta_l}}

\newcommand{\hpaai}{\hat{P}^{a}_{\alpha_i}}
\newcommand{\hpaaj}{\hat{P}^{a}_{\alpha_j}}
\newcommand{\hpaak}{\hat{P}^{a}_{\alpha_k}}

\newcommand{\Paa}{{P}^{a}_{\alpha}}
\newcommand{\Paae}{{P}^{a}_{\alpha_1}}
\newcommand{\Paat}{{P}^{a}_{\alpha_2}}
\newcommand{\Paas}{{P}^{a}_{\alpha_3}}
\newcommand{\Paai}{{P}^{a}_{\alpha_i}}
\newcommand{\Paaj}{{P}^{a}_{\alpha_j}}
\newcommand{\Paak}{{P}^{a}_{\alpha_k}}

\newcommand{\paai}{{p}^{a}_{\alpha_i}}
\newcommand{\paaj}{{p}^{a}_{\alpha_j}}
\newcommand{\paak}{{p}^{a}_{\alpha_k}}
\newcommand{\Pbb}{{P}^{b}_\beta}
\newcommand{\hpnae}{{\hat{P}^{a\bot}_{\alpha_1}}}
\newcommand{\hpnat}{{\hat{P}^{a\bot}_{\alpha_2}}}
\newcommand{\hpnas}{{\hat{P}^{a\bot}_{\alpha_3}}}

\newcommand{\ebb}{e^b_{\beta}}
\newcommand{\eaa}{e^a_{\alpha}}
\newcommand{\ebbl}{e^b_{\beta_l}}

\newcommand{\ebbe}{e^b_{\beta_1}}
\newcommand{\ebbt}{e^b_{\beta_2}}
\newcommand{\ebbs}{e^b_{\beta_3}}
\newcommand{\eaai}{e^a_{\alpha_i}}
\newcommand{\eaaj}{e^a_{\alpha_j}}
\newcommand{\eaak}{e^a_{\alpha_k}}

\newcommand{\eaam}{e^a_{\alpha_m}}

\newcommand{\bra}[1]{\left\langle{#1}\right\vert}
\newcommand{\ket}[1]{\left\vert{#1}\right\rangle}

\newcommand{\w}{w}
\newcommand{\cw}{\overline{w}}
\newcommand{\gam}[1]{\gamma_{#1}}

\title{Quantum-like representation algorithm for trichotomous observables}
\author{Peter Nyman and Irina Basieva\footnote{Research fellowship of Swedish Institute.}\\
\\ International Center for Mathematical Modelling
\\in Physics and Cognitive Sciences,\\
Linnaeus University, S-35195, Sweden}

\begin{document}
\maketitle
\begin{abstract}
We study the problem of representation of statistical data (of any origin) by a complex probability amplitude.
This paper is devoted to representation of data collected from measurements of two trichotomous observables. The complexity of the problem eventually increased comparing to with the case of dichotomous observables.  We see that only special statistical data (satisfying s number of nonlinear constraints) have the quantum--like representation. 
\end{abstract}
\section{Introduction}

The problem of inter-relation between classical and quantum probabilistic data was discussed in numerous papers 
(from various points of view), see, e.g., \cite{VN,GD1,GD2,GD3,DV,BL,NAN,NAN1,AL,K6}. We are interested in the problem of representation of probabilistic 
data of any origin\footnote{Thus it need not be produced by quantum measurements; it can be collected in e.g. psychology,
see \cite{K9}.} by complex probability amplitude, so to say a ``wave function''. This problem was discussed in very detail
in \cite{K8}. It has two sources. One is purely mathematical: to describe a class of data which permits the quantum-like (QL)
representation. Another reason to create the QL-representation of data is not so straightforward as the first one.
In \cite{K9}  A. Khrennikov presented a hypothesis that biological systems
might use complex probabilistic amplitudes (``mental wave functions'')  in processing of statistical data.
If this hypothesis is correct, then these   amplitudes can be reconstructed on the basis of 
collected experimental data. In psychology this approach got the name ``constructive wave function approach''.

A general QL-representation algorithm (QLRA) was presented in \cite{K8}. This algorithm is based 
on the {\it  formula of total probability with interference term} -- a disturbance of the standard formula of total 
probability. Starting with experimental probabilistic data, QLRA produces a complex probability amplitude such that 
probability can be reconstructed by using Born's rule. 

Although the formal scheme of QLRA works for multi-valued observables of an arbitrary dimension, the description of the class of 
probabilistic data which can be transfered into QL-amplitudes (the domain of application of QLRA)  
depends very much on the dimension. In \cite{N1} the simplest case of data generated by dichotomous observables was studied.
In this paper we study trichotomous observables. The complexity of the problem increases incredibly comparing with the two
dimensional case.    

Finally, we remark that our study is closely related to the triple slit interference experiment and 
Sorkin's equality \cite{Urbasi}. This experiment provides an important test of foundations of QM.

The scheme of presentation is the following one. We start with observables given by QM and derive 
constraints on phases which are necessary and sufficient for the QL-representation. Then we use 
these constraints to produce complex amplitudes from data (of any origin); some examples, including numerical, are 
given.      

\section{Trichotomous incompatible quantum observables}

\subsection{Probabilities}
\label{HH7}

Let $\hat{a}$ and $\hat{b}$ be two self-adjoint operators in three dimensional complex Hilbert space representing two trichotomous incompatible observables $a$ and $b$. They take values $a=\alpha_i,\;i=1,2,3$ and $b=\beta_l, l=1,2,3$ --
spectra of operators. We assume that the operators have nondegenerate spectra, i.e., $\alpha_i \not=\alpha_j, \beta_i \not=\beta_j, i \not=j.$
Consider corresponding eigenvectors: $$\hat{a} e^{a}_{\alpha_i}={\alpha_i} e^{a}_{\alpha_i}, \: \hat{b} e^{b}_{\beta_l}={\beta_l} e^{b}_{\beta_l}.$$
Denote by $\hpaai=\ket{\eaai}\bra{\eaai}$ and $\hpbbl=\vert{\ebbl}\rangle\langle{\ebbl}\vert$ one dimensional projection operators and by $\Paa$ and $\Pbb$ the observables repressed by there projections. Consider also projections \begin{align}\hpnae&=\hpaat+\hpaas,\\\nonumber \hpnat&=\hpaae+\hpaas,\\\nonumber \hpnas&=\hpaae+\hpaat.\end{align}
Here the observable $\Paai=1$ if the result of the $a$-measurement is $a=\alpha_i$ and $\Paai=0$ if  $a\not=\alpha_i.$
The  observables $\Pbb$ are defined in the same way.
We have the following relation between  events corresponding to measurements
$
[\Paae=0]=[\Paat=1]\vee[\Paas=1], \; [\Paat=0]=[\Paae=1]\vee[\Paas=1],
[\Paas=0]=[\Paae=1]\vee[\Paat=1].
$
There are given (by the QM-formalism)  the probabilities 
\begin{align}\label{prob1} p^b_{\beta}\equiv P_\psi (b=\beta)= ||\hpbb \psi||^2=|\langle{\psi} \vert{\ebb}\rangle|^2,
\\\nonumber p^a_{\alpha}\equiv P_\psi (a=\alpha)=||\hpaa \psi||^2=|\langle{\psi}\vert {\eaa}\rangle|^2.
\end{align}
There are also given (by the QM-formalism) conditional (transition) probabilities 
\begin{equation}
p^{b|a}_{\beta\alpha} \equiv  P_\psi(b=\beta|\Paa=1)=||\hpbb \hpaa \psi||^2/||\hpaa\psi||^2=|\langle\eaa|\ebb\rangle|^2.
\end{equation} 
We remark that non degeneration of the spectra implies that they do not depend on $\psi.$  
Moreover, the matrix of transition probabilities is {\it doubly stochastic}.
There are also given ($\psi$-depend) probabilities 
$$
p_{\beta_l\alpha_{kj}}^{b|a} \equiv P_\psi(b=\beta_l|\Paak = 1\vee \Paaj = 1)= P_\psi(b=\beta_l|\Paai=0) =\frac{\left\vert\right\vert\hpbbl(\hpaaj+\hpaak)\psi\left\vert\right\vert^2}{\left\vert\right\vert
(\hpaaj+\hpaak)\psi\left\vert\right\vert^2},
$$ 
where $\;ij,k\in \{1,2,3\},\; j,k  \neq i.$
We have 
$$
\frac{\left\vert\right\vert\hpbb(\hpaaj+\hpaak)\psi\left\vert\right\vert^2}{\left\vert\right\vert(\hpaaj+\hpaak)\psi\left\vert\right\vert^2}=
\frac{\left\vert\right\vert\vert{\ebb}\rangle\langle\ebb\vert{\eaaj}\rangle\langle{\eaaj}\vert\psi\rangle+\vert{\ebb}\rangle\langle\ebb\vert{\eaak}\rangle\langle{\eaak}\vert\psi\rangle\left\vert\right\vert^2}{\left\vert\right\vert\vert{\eaaj}\rangle\langle{\eaaj}\vert\psi\rangle+\vert{\eaak}\rangle\langle{\eaak}\vert\psi\rangle\left\vert\right\vert^2}
$$
$$
=\frac{\left\vert\right\vert\vert{\ebb}\rangle\left\vert\right\vert^2\vert\langle\ebb\vert{\eaaj}\rangle\langle{\eaaj}\vert\psi\rangle+\langle\ebb\vert{\eaak}\rangle\langle{\eaak}\vert\psi\rangle\vert^2}{\vert\langle{\eaaj}\vert\psi\rangle\vert^2+\vert\langle{\eaak}\vert\psi\rangle\vert^2}
=\frac{\vert\langle\ebb\vert{\eaaj}\rangle\langle{\eaaj}\vert\psi\rangle+\langle\ebb\vert{\eaak}\rangle\langle{\eaak}\vert\psi\rangle\vert^2}{\vert\langle{\eaaj}\vert\psi\rangle\vert^2+\vert\langle{\eaak}\vert\psi\rangle\vert^2}.
$$ 
Note that $p_{\beta_l\alpha_{kj}}^{b|a}=p_{\beta_l\alpha_{jk}}^{b|a}$.
\subsection{Probability amplitudes}

Set 
$
\psi_{\beta} = \langle\psi |\ebb\rangle.
$
Then by Born's rule 
\begin{equation}
\label{RRTT}
p_\beta^b = \vert \psi_\beta\vert^2.
\end{equation}
We have 
\begin{equation}
\label{RRTT1}
\psi = \sum_\beta \psi_\beta \ebb.
\end{equation}
Thus these amplitudes give a possibility to reconstruct the state.
We remark that $\psi= \sum_\alpha \hpaa \psi,$ and, hence,
\begin{equation}
\label{RRTT2}
\psi_{\beta}= \sum_\alpha \langle \hpaa \psi |\ebb\rangle.
\end{equation}
Each amplitude $\psi_{\beta}$ can be represented as the sum of three subamplitudes
\begin{equation}
\label{RRTT3}
\psi_{\beta}= \sum_\alpha \psi_{\beta\alpha}
\end{equation}
 given by 
\begin{equation}\label{RRTT4}\psi_{\beta\alpha}=\langle\hpaa\psi|\ebb\rangle=\langle\psi|\eaa\rangle\langle\eaa|\ebb\rangle.
\end{equation} 
Hence, one can reconstruct the state $\psi$ on the basis of nine amplitudes $\psi_{\beta\alpha}.$
We remark that
$|\psi_{\beta_l\alpha_i}|^2=|\langle\psi|\eaai\rangle\langle\eaai|\ebbl\rangle|^2=p^a_{\alpha_i} p_{\beta_l\alpha_i}^{b|a}.
$
In this notations
\begin{equation}\label{pblakj} p_{\beta_l\alpha_{kj}}^{b|a}=|\psi_{\beta_l\alpha_k}+\psi_{\beta_l\alpha_j}|^2/(p^a_{\alpha_j}+p^a_{\alpha_k}).
\end{equation}
Here $|\psi_{\beta_l\alpha_i}|=\sqrt{p^{a}_{\alpha_i}p^{b|a}_{\beta_l,\alpha_i}}$ 
and therefore  $$\psi_{\beta_l\alpha_i}=\sqrt{p^{a}_{\alpha_i}p^{b|a}_{\beta_l,\alpha_i}}e^{i \varphi_{\beta_l\alpha_i}},$$ where $\varphi_{\beta_l\alpha_i}=\arg \psi_{\beta_l\alpha_i}.$
Moreover, 
\begin{equation}
\label{tio}
\langle \psi|\eaam\rangle
=\sqrt{p_{\alpha_m}^a}e^{i \xi_{\alpha_i}}, \; \langle \eaam|\ebbl\rangle =\sqrt{p_{\beta_l \alpha_m}^{ab}}e^{i \theta_{\beta_l\alpha_m}}.
\end{equation}
Hence,
\begin{equation}\label{tioH}
\varphi_{\beta_l \alpha_m}=\theta_{\beta_l \alpha_m}+ \xi_{\alpha_i}.
\end{equation}

We have a system of equations for phases $\psi_{\beta_l\alpha_i}$  for $i,j,k,l \in\{1,2,3\}$,
\begin{eqnarray}
|\psi_{\beta_l\alpha_i}+\psi_{\beta_l\alpha_j}|^2&=&|\sqrt{p^{a}_{\alpha_i}p^{b|a}_{\beta_l,\alpha_i}}e^{i\varphi_{\beta_l\alpha_i}}+\sqrt{p^{a}_{\alpha_i}p^{b|a}_{\beta_l,\alpha_j}}e^{i\varphi_{\beta_l\alpha_j}}|^2\\\nonumber&=&p^{a}_{\alpha_i}p^{b|a}_{\beta_l,\alpha_i}+p^{a}_{\alpha_j}p^{b|a}_{\beta_l,\alpha_j}\\\nonumber &+&2\cos(\varphi_{\beta_l\alpha_i}-\varphi_{\beta_l\alpha_j})\sqrt{p^{a}_{\alpha_i}p^{b|a}_{\beta_l,\alpha_i}p^{a}_{\alpha_j}p^{b|a}_{\beta_l,\alpha_j}}.
\end{eqnarray}

We set
\begin{equation}\label{sju}
\lambda_{l,ij}\equiv\cos(\varphi_{\beta_l\alpha_i}-\varphi_{\beta_l\alpha_j})
\end{equation}
and we have
\begin{equation}\label{sju7}
\lambda_{l,ij} =\frac{(\paai+\paaj)p_{\beta_l\alpha_{ij}}^{b|a}-(\paai p_{\beta_l\alpha_{i}}^{b|a}+\paaj p_{\beta_l\alpha_{j}}^{b|a})}{2\sqrt{\paai p_{\beta_l\alpha_{i}}^{b|a}\paaj p_{\beta_l\alpha_{j}}^{b|a}}}.
\end{equation}

We call $\lambda_{l,ij}$ for the coefficients of interference.
\subsection{Formula of total probability with interference term}

By using the definition of the amplitude $\psi_{\beta_l\alpha_i}=\langle\psi|\eaa\rangle\langle\eaa|\ebb\rangle$ we obtain
\begin{equation}
\label{atta} 
p_{\beta_l}^b= |\psi_{\beta_l\alpha_i}+\psi_{\beta_l\alpha_j}+\psi_{\beta_l\alpha_k}|^2=\vert \langle\psi|\eaai\rangle\langle\eaai|\ebbl\rangle+\langle\psi|\eaaj\rangle\langle\eaaj|\ebbl\rangle
+\langle\psi|\eaak\rangle\langle\eaak|\ebbl\rangle|^2.
\end{equation}
$$
=|\langle\psi|\eaai\rangle\langle\eaai|\ebbl\rangle|^2+|\langle\psi|\eaaj\rangle\langle\eaaj|\ebbl\rangle|^2
+|\langle\psi|\eaak\rangle\langle\eaak|\ebbl\rangle|^2+
$$
$$
+ \langle\psi|\eaai\rangle \langle\eaai|\ebbl\rangle \langle\ebbl|\eaaj\rangle \langle\eaaj|\psi\rangle
+ \langle\psi|\eaai\rangle \langle\eaai|\ebbl\rangle \langle\ebbl|\eaak\rangle \langle\eaak|\psi\rangle
$$
$$
+ \langle\psi|\eaaj\rangle \langle\eaaj|\ebbl\rangle \langle\ebbl|\eaak\rangle \langle\eaak|\psi\rangle
+\langle\eaai|\psi\rangle \langle\ebbl|\eaai\rangle \langle\psi|\eaaj\rangle\langle\eaaj|\ebbl\rangle
$$
$$
+\langle\eaai|\psi\rangle \langle\ebbl|\eaai\rangle \langle\psi|\eaak\rangle\langle\eaak|\ebbl\rangle
+\langle\eaaj|\psi\rangle \langle\ebbl|\eaaj\rangle \langle\psi|\eaak\rangle\langle\eaak|\ebbl\rangle .
$$
Finally, we obtain  
\begin{align}\label{elva} p_{\beta_l}^b
&=p_{\alpha_i}^a p_{\beta_l \alpha_i}^{b|a}+p_{\alpha_j}^a p_{\beta_l \alpha_j}^{b|a}+p_{\alpha_k}^a p_{\beta_l \alpha_k}^{b|a}
\\\nonumber&+2\cos(\varphi_{\beta_l \alpha_i}-\varphi_{\beta_l \alpha_j})\sqrt{p_{\alpha_i}p_{\alpha_j}p_{\beta_l \alpha_i}^{b|a}p_{\beta_l \alpha_j}^{b|a}}
+2\cos(\varphi_{\beta_l \alpha_i}-\varphi_{\beta_l \alpha_k})\sqrt{p_{\alpha_i}p_{\alpha_k}p_{\beta_l \alpha_i}^{b|a}p_{\beta_l \alpha_k}^{b|a}}
\\\nonumber&+2\cos(\varphi_{\beta_l \alpha_j}-\varphi_{\beta_l \alpha_k})\sqrt{p_{\alpha_j}p_{\alpha_k}p_{\beta_l \alpha_j}^{b|a}p_{\beta_l \alpha_k}^{b|a}}.
\end{align}
This is nothing else than the formula of total probability with the interference term. It can be considered \cite{K10}  as a perturbation 
of the classical formula of total probability
\begin{equation} 
\label{elva1} 
p_{\beta_l}^b =p_{\alpha_i}^a p_{\beta_l \alpha_i}^{b|a}+p_{\alpha_j}^a p_{\beta_l \alpha_j}^{b|a}+p_{\alpha_k}^a p_{\beta_l \alpha_k}^{b|a}.
\end{equation}
If all coefficients of interferes $\lambda_{l.ij}=0$, then \eqref{elva} coincides with \eqref{elva1}.
\subsection{Sorkin's equality in conditional probabilistic form}

We will derive Sorkin's equality by putting \eqref{sju} in \eqref{elva},  
\begin{align}
p_{\beta_l}^b&=p^{a}_{\alpha_i}p^{b|a}_{\beta_l\alpha_i}+p^{a}_{\alpha_j}p^{b|a}_{\beta_l\alpha_j}+p^{a}_{\alpha_k}p^{b|a}_{\beta_l\alpha_k} 
+\left((\paai+\paaj)p^{b|a}_{\beta_l\alpha_{i\vee j}}-(\paai p^{b|a}_{\beta_l\alpha_i}+\paaj p^{b|a}_{\beta_l\alpha_j})\right) 
\\\nonumber&+\left((\paai+\paak)p^{b|a}_{\beta_l\alpha_{i\vee k}}-(\paai p^{b|a}_{\beta_l\alpha_i}+\paak p^{b|a}_{\beta_l\alpha_k})\right) 
\\\nonumber&
+\left((\paaj+\paak)p^{b|a}_{\beta_l\alpha_{j\vee k}}-(\paaj p^{b|a}_{\beta_l\alpha_j}+\paak p^{b|a}_{\beta_l\alpha_k})\right)
\\\nonumber&=(\paai+\paaj)p^{b|a}_{\beta_l\alpha_{i\vee j}}+(\paai+\paak)p^{b|a}_{\beta_l\alpha_{i\vee k}}+(\paaj+\paak)p^{b|a}_{\beta_l\alpha_{j\vee k}}
\\\nonumber&-\left(\paai p^{b|a}_{\beta_l\alpha_i}+\paaj p^{b|a}_{\beta_l\alpha_j}+\paak p^{b|a}_{\beta_l\alpha_k}\right)
\\\nonumber&=\paai(p^{b|a}_{\beta_l\alpha_{i\vee j}}+p^{b|a}_{\beta_l\alpha_{i\vee k}}-p^{b|a}_{\beta_l\alpha_i})
+\paaj(p^{b|a}_{\beta_l\alpha_{i\vee j}}+p^{b|a}_{\beta_l\alpha_{j\vee k}}-p^{b|a}_{\beta_l\alpha_j})
\\\nonumber&+ \paak(p^{b|a}_{\beta_l\alpha_{i\vee k}}+p^{b|a}_{\beta_l\alpha_{j\vee k}}-p^{b|a}_{\beta_l\alpha_k}).
\end{align}
This gives us the following constraint on the probabilities
\begin{equation}
\label{CBORN}
p^{b}_{\beta_l}= \paai(p^{b|a}_{\beta_l\alpha_{ij}}+p^{b|a}_{\beta_l\alpha_{ik}}-p^{b|a}_{\beta_l\alpha_i})
\end{equation}
$$
+\paaj(p^{b|a}_{\beta_l\alpha_{ij}}+p^{b|a}_{\beta_l\alpha_{jk}}-p^{b|a}_{\beta_l\alpha_j})
+\paak(p^{b|a}_{\beta_l\alpha_{ik}}+p^{b|a}_{\beta_l\alpha_{jk}}-p^{b|a}_{\beta_l\alpha_k}).
$$
This equation coupling various quantum probabilities can be considered as incrypting of Born's rule 
by using the language of probabilities. This is the discrete version of famous Sorkin equality \cite{S1, S2}.

\subsection{Unitarity of transition operator}
\label{Un}

We now remark that bases consisting of $\hat{a}$- and $\hat{b}$-eigenvectors are orthogonal; hence the operator of transition from
one basis to another is unitarity. We can always select the $b$-basis in the canonical way 
\begin{equation}
\label{L1}
e^b_{\beta_1}=\left(
\begin{array}{ll}
 1\\
 0\\
 0
 \end{array}
 \right ), 
 \; e^b_{\beta_2}=\left(
\begin{array}{ll}
 0\\
 1\\
 0
 \end{array}
 \right ),\; 
e^b_{\beta_3}=\left(
\begin{array}{ll}
 0\\
 0\\
 1
 \end{array}
 \right ).
\end{equation}
In this system of coordinates the $a$-basis has the form
\begin{equation}
\label{L2}
e^a_{\alpha_1}=\left(
\begin{array}{ll}
 \sqrt{p_{\beta_1 \alpha_1}}e^{i \theta_{\beta_1 \alpha_1}}\\
 \sqrt{p_{\beta_2 \alpha_1}}e^{i \theta_{\beta_2 \alpha_1}}\\
 \sqrt{p_{\beta_3 \alpha_1}}e^{i \theta_{\beta_3 \alpha_1}}
 \end{array}
 \right ), 
\; 
e^a_{\alpha_2}=\left(
\begin{array}{ll}
 \sqrt{p_{\beta_1 \alpha_2}}e^{i \theta_{\beta_1 \alpha_2}}\\
\sqrt{p_{\beta_2 \alpha_2}}e^{i \theta_{\beta_2 \alpha_2}} \\
 \sqrt{p_{\beta_3 \alpha_2}}e^{i \theta_{\beta_3 \alpha_2}}
 \end{array}
 \right ), 
\; 
e^a_{\alpha_3}=\left(
\begin{array}{ll}
\sqrt{p_{\beta_1 \alpha_3}}e^{i \theta_{\beta_1 \alpha_3}}\\
\sqrt{p_{\beta_3 \alpha_2}}e^{i \theta_{\beta_3 \alpha_2}}\\
 \sqrt{p_{\beta_3 \alpha_3}}e^{i \theta_{\beta_3 \alpha_3}}
 \end{array}
 \right ) 
\end{equation}
The matrix
$$
U =\left(
\begin{array}{ll}
 \sqrt{p_{\beta_1 \alpha_1}} e^{i \theta_{\beta_1 \alpha_1}} \; \sqrt{p_{\beta_1 \alpha_2}} e^{i \theta_{\beta_1 \alpha_2}}\; 
\sqrt{p_{\beta_1 \alpha_3}} e^{i \theta_{\beta_1 \alpha_3}} \\
 \sqrt{p_{\beta_2 \alpha_1}} e^{i \theta_{\beta_2 \alpha_1}}\; \sqrt{p_{\beta_2 \alpha_2}} e^{i \theta_{\beta_2 \alpha_2}}\; \sqrt{p_{\beta_2 \alpha_3}} e^{i \theta_{\beta_2 \alpha_3}} \\
 \sqrt{p_{\beta_3 \alpha_1}} e^{i \theta_{\beta_3 \alpha_1}}\; \sqrt{p_{\beta_3 \alpha_2}} e^{i \theta_{\beta_3 \alpha_2}}\; 
 \sqrt{p_{\beta_3 \alpha_3}} e^{i \theta_{\beta_3 \alpha_3}}
 \end{array}
 \right)
$$ 
is unitary. Hence, we have the system of equations 
\begin{equation}
\label{UIZ}
 \sum_m \sqrt{p_{\beta_m \alpha_i}p_{\beta_m \alpha_j}} e^{i (\theta_{\beta_m \alpha_i} -\theta_{\beta_m \alpha_j)}}= 0
 \end{equation}
 or 
\begin{equation}
\label{UI1}
 \sum_m \sqrt{p_{\beta_m \alpha_i}p_{\beta_m \alpha_j}} \cos(\theta_{\beta_m \alpha_i} -\theta_{\beta_m \alpha_j})=0,
 \end{equation}
\begin{equation}
\label{UI2}
 \sum_m \sqrt{p_{\beta_m \alpha_i}p_{\beta_m \alpha_j}} \sin(\theta_{\beta_m \alpha_i} -\theta_{\beta_m \alpha_j})= 0,
 \end{equation}
 where $(i\not=j)$ (For $i=j,$ the unitarity condition is equivalent to normalization 
of the sum of probabilities by one)
+.
We now recall that the phases of the basis vectors $e_\alpha^a$ are coupled with the phases of the amplitudes 
$\psi_{\beta \alpha}$ by (\ref{tioH}). Hence, we obtain a system of constraints on the later phases
\begin{equation}
\label{UI1J}
 \sum_m \sqrt{p_{\beta_m \alpha_i}p_{\beta_m \alpha_j}} \cos[(\phi_{\beta_m \alpha_i} -\phi_{\beta_m \alpha_j})+ (\xi_{\alpha_j}- \xi_{\alpha_i})]= 0,
 \end{equation}
\begin{equation}
\label{UI2J}
 \sum_m \sqrt{p_{\beta_m \alpha_i}p_{\beta_m \alpha_j}} \sin[(\phi_{\beta_m \alpha_i} -\phi_{\beta_m \alpha_j})+ (\xi_{\alpha_j}- \xi_{\alpha_i})]=0.
\end{equation}
Thus 
\begin{equation}
\label{UI1M}
 \cos(\xi_{\alpha_j}- \xi_{\alpha_i}) \sum_m \sqrt{p_{\beta_m \alpha_i}p_{\beta_m \alpha_j}} \cos(\phi_{\beta_m \alpha_i} -\phi_{\beta_m \alpha_j})
\end{equation}
$$
 - \sin(\xi_{\alpha_j}- \xi_{\alpha_i}) \sum_m \sqrt{p_{\beta_m \alpha_i}p_{\beta_m \alpha_j}} \sin(\phi_{\beta_m \alpha_i} -\phi_{\beta_m \alpha_j})= 0;
$$
\begin{equation}
\label{UI1M1}
 \cos(\xi_{\alpha_j}- \xi_{\alpha_i}) \sum_m \sqrt{p_{\beta_m \alpha_i}p_{\beta_m \alpha_j}} \sin(\phi_{\beta_m \alpha_i} -\phi_{\beta_m \alpha_j})
\end{equation}
$$
 + \sin(\xi_{\alpha_j}- \xi_{\alpha_i}) \sum_m \sqrt{p_{\beta_m \alpha_i}p_{\beta_m \alpha_j}} \cos(\phi_{\beta_m \alpha_i} -\phi_{\beta_m \alpha_j})= 0.
$$
Suppose now the following equalities hold 
\begin{equation}
\label{UI1H}
 \sum_m \sqrt{p_{\beta_m \alpha_i}p_{\beta_m \alpha_j}} \cos(\phi_{\beta_m \alpha_i} -\phi_{\beta_m \alpha_j})=0,
 \end{equation}
\begin{equation}
\label{UI2H}
 \sum_m \sqrt{p_{\beta_m \alpha_i}p_{\beta_m \alpha_j}} \sin(\phi_{\beta_m \alpha_i} -\phi_{\beta_m \alpha_j})= 0.
 \end{equation}
Then by (\ref{UI1M}),  (\ref{UI1M1}) the equalities (\ref{UI1}), (\ref{UI2}) and, hence, (\ref{UIZ}) hold. Thus 
conditions (\ref{UI1H}),  (\ref{UI2H}) imply unitarity of $U.$  It is clear that in turn (\ref{UI1}), (\ref{UI2})
imply  (\ref{UI1H}),  (\ref{UI2H}) for arbitrary $\xi_\alpha.$ Thus the later conditions are equivalent to unitarity 
of $U.$

\section{Mutually unbiased bases}

In previous considerations we introduced the coefficients of interference $\lambda_{l,ij}$ on the basis of the phases $\varphi_{\beta_l \alpha_i},\;\varphi_{\beta_l \alpha_j}$ by $\lambda_{l,ij}=\cos(\varphi_{\beta_l \alpha_i}-\varphi_{\beta_l \alpha_j})$. However, we see that they also could be defined by using only probabilistic data, see \eqref{sju7}.
Can we the go other way around and to find phases $\varphi_{\beta_l \alpha_i},\;\varphi_{\beta_l \alpha_j}$ on the basis of the interference coefficients given by \eqref{sju7}? We will study this general problem in section \ref{cbr}.
Now we would like consider an example. To be sure that the problem has a solution, we start with data and the corresponding interference coefficients generated by QM. In section \ref{cbr} we shall operate with statistical data an arbitrary origin.  
First we show that there exist probabilistic data such that \eqref{UI1H} and \eqref{UI2H} hold, therefore consider the following situation. Let $p_{\beta_i \alpha_j}=1/3$ where $i,j=1,2,3 $.
We will also put 
\begin{equation}
\theta_{\beta_2 \alpha_1}=\theta_{\beta_3 \alpha_3}=-\theta_{\beta_2 \alpha_2}=-\theta_{\beta_3 \alpha_2}=2\pi/3,
 \end{equation}
 where all the other $\theta_{\beta_i \alpha_j}=0, \quad i,j=1,2,3$.
 We insert this in the basis in \eqref{L2} and obtain the orthonormal basis
\begin{equation}
 \label{Mub1}
e^a_{\alpha_1}=\frac{1}{\sqrt{3}}\left(
\begin{array}{l}1\\ \w \\1 \end{array}
 \right), 
\; 
e^a_{\alpha_2}=\frac{1}{\sqrt{3}}\left(
\begin{array}{l}
 1\\
\cw \\
\cw
 \end{array}
 \right), 
\; 
e^a_{\alpha_3}=\frac{1}{\sqrt{3}}\left(
\begin{array}{l}
1\\
1\\
\w
 \end{array}
 \right), 
\end{equation}
where $w = e^{i 2\pi/3}$. Bases \eqref{L1} and \eqref{Mub1}  are mutually unbiased, i.e. ,$|\langle e^b_{\beta_j}|\eaai\rangle|=1/3.$
Then let $$\psi=\frac{1}{\sqrt{3}}(e_{\beta_1}+e^{i \gam{1}}e_{\beta_2}+e^{i \gam{2}}e_{\beta_3}).$$ Note that $| \psi|^2 = 1$. Then by equation \eqref{prob1} and after some calculations this gives that,
\begin{equation}
\label{Palpha1}
p_{\alpha_1}=| \langle{\ebbe}\vert{\psi}\rangle|^2=\frac{1}{9} \left(3-\cos\left(\gamma _1\right)-\cos\left(\gamma _1-\gamma _2\right)+2 \cos\left(\gamma _2\right)+\sqrt{3} \sin\left(\gamma _1\right)+\sqrt{3} \sin\left(\gamma _1-\gamma _2\right)\right),
\end{equation}
\begin{equation}
\label{Palpha2}
p_{\alpha_2}=| \langle{\ebbt}\vert{\psi}\rangle|^2=\frac{1}{9} \left(3-\cos\left(\gamma _1\right)+2 \cos\left(\gamma _1-\gamma _2\right)-\cos\left(\gamma _2\right)-\sqrt{3} \sin\left(\gamma _1\right)-\sqrt{3} \sin\left(\gamma _2\right)\right),
\end{equation}
\begin{equation}
\label{Palpha3}
p_{\alpha_3}=|\langle {\ebbs}\vert{\psi}\rangle|^2=\frac{1}{9} \left(3+2 \cos\left(\gamma _1\right)-\cos\left(\gamma _1-\gamma _2\right)-\cos\left(\gamma _2\right)-\sqrt{3} \sin\left(\gamma _1-\gamma _2\right)+\sqrt{3} \sin\left(\gamma _2\right)\right).
\end{equation}
It is straightforward to see that $p_{\alpha_1}+p_{\alpha_2}+p_{\alpha_3}=1$.
The conditional probabilities\footnote{Recall that $p_{\beta_l\alpha_{kj}}^{b|a}=p_{\beta_l\alpha_{jk}}^{b|a}$.} $p_{\beta_l\alpha_{kj}}$ for $ l,k,j=1,2,3,\;\;k\neq j$ are calculated by \eqref{pblakj} 
\begin{align}
p_{\beta_1\alpha_{12}}&=-\frac{4 \cos \left(\gamma _1\right)+\cos \left(\gamma _1-\gamma
   _2\right)-2 \cos \left(\gamma _2\right)+\sqrt{3} \sin \left(\gamma
   _1-\gamma _2\right)+2 \sqrt{3} \sin \left(\gamma _2\right)-6}{3
   \left(-2 \cos \left(\gamma _1\right)+\cos \left(\gamma _1-\gamma
   _2\right)+\cos \left(\gamma _2\right)+\sqrt{3} \sin \left(\gamma
   _1-\gamma _2\right)-\sqrt{3} \sin \left(\gamma _2\right)+6\right)}\\\nonumber
p_{\beta_1\alpha_{13}}&=\frac{2 \left(\cos \left(\gamma _1\right)+\cos \left(\gamma _1-\gamma
   _2\right)+\cos \left(\gamma _2\right)+\sqrt{3} \sin \left(\gamma
   _1\right)+\sqrt{3} \sin \left(\gamma _2\right)+3\right)}{3
   \left(\cos \left(\gamma _1\right)-2 \cos \left(\gamma _1-\gamma
   _2\right)+\cos \left(\gamma _2\right)+\sqrt{3} \sin \left(\gamma
   _1\right)+\sqrt{3} \sin \left(\gamma _2\right)+6\right)}\\\nonumber
p_{\beta_1\alpha_{23}}&=-\frac{-2 \cos \left(\gamma _1\right)+\cos \left(\gamma _1-\gamma
   _2\right)+4 \cos \left(\gamma _2\right)+2 \sqrt{3} \sin \left(\gamma
   _1\right)-\sqrt{3} \sin \left(\gamma _1-\gamma _2\right)-6}{3
   \left(\cos \left(\gamma _1\right)+\cos \left(\gamma _1-\gamma
   _2\right)-2 \cos \left(\gamma _2\right)-\sqrt{3} \sin \left(\gamma
   _1\right)-\sqrt{3} \sin \left(\gamma _1-\gamma _2\right)+6\right)}\\\nonumber
p_{\beta_2\alpha_{12}}&=\frac{-4 \cos \left(\gamma _1\right)+2 \cos \left(\gamma _1-\gamma
   _2\right)-\cos \left(\gamma _2\right)+2 \sqrt{3} \sin \left(\gamma
   _1-\gamma _2\right)+\sqrt{3} \sin \left(\gamma _2\right)+6}{3 \left(-2
   \cos \left(\gamma _1\right)+\cos \left(\gamma _1-\gamma _2\right)+\cos
   \left(\gamma _2\right)+\sqrt{3} \sin \left(\gamma _1-\gamma
   _2\right)-\sqrt{3} \sin \left(\gamma _2\right)+6\right)}\\\nonumber
p_{\beta_2\alpha_{13}}&=-\frac{-2 \cos \left(\gamma _1\right)+4 \cos \left(\gamma _1-\gamma
   _2\right)+\cos \left(\gamma _2\right)-2 \sqrt{3} \sin \left(\gamma
   _1\right)+\sqrt{3} \sin \left(\gamma _2\right)-6}{3 \left(\cos
   \left(\gamma _1\right)-2 \cos \left(\gamma _1-\gamma _2\right)+\cos
   \left(\gamma _2\right)+\sqrt{3} \sin \left(\gamma _1\right)+\sqrt{3}
   \sin \left(\gamma _2\right)+6\right)}\\\nonumber
p_{\beta_2\alpha_{23}}&=\frac{2 \left(\cos \left(\gamma _1\right)+\cos \left(\gamma _1-\gamma
   _2\right)+\cos \left(\gamma _2\right)-\sqrt{3} \sin \left(\gamma
   _1\right)-\sqrt{3} \sin \left(\gamma _1-\gamma _2\right)+3\right)}{3
   \left(\cos \left(\gamma _1\right)+\cos \left(\gamma _1-\gamma
   _2\right)-2 \cos \left(\gamma _2\right)-\sqrt{3} \sin \left(\gamma
   _1\right)-\sqrt{3} \sin \left(\gamma _1-\gamma _2\right)+6\right)}\\\nonumber
p_{\beta_3\alpha_{12}}&=\frac{2 \left(\cos \left(\gamma _1\right)+\cos \left(\gamma _1-\gamma
   _2\right)+\cos \left(\gamma _2\right)+\sqrt{3} \sin \left(\gamma
   _1-\gamma _2\right)-\sqrt{3} \sin \left(\gamma _2\right)+3\right)}{3
   \left(-2 \cos \left(\gamma _1\right)+\cos \left(\gamma _1-\gamma
   _2\right)+\cos \left(\gamma _2\right)+\sqrt{3} \sin \left(\gamma
   _1-\gamma _2\right)-\sqrt{3} \sin \left(\gamma _2\right)+6\right)}\\\nonumber
p_{\beta_3\alpha_{13}}&=-\frac{\cos \left(\gamma _1\right)+4 \cos \left(\gamma _1-\gamma
   _2\right)+\sqrt{3} \sin \left(\gamma _1\right)-2 \left(\cos
   \left(\gamma _2\right)+\sqrt{3} \sin \left(\gamma _2\right)+3\right)}{3
   \left(\cos \left(\gamma _1\right)-2 \cos \left(\gamma _1-\gamma
   _2\right)+\cos \left(\gamma _2\right)+\sqrt{3} \sin \left(\gamma
   _1\right)+\sqrt{3} \sin \left(\gamma _2\right)+6\right)}\\\nonumber
p_{\beta_3\alpha_{23}}&=-\frac{\cos \left(\gamma _1\right)-2 \cos \left(\gamma _1-\gamma
   _2\right)+4 \cos \left(\gamma _2\right)-\sqrt{3} \sin \left(\gamma
   _1\right)+2 \sqrt{3} \sin \left(\gamma _1-\gamma _2\right)-6}{3
   \left(\cos \left(\gamma _1\right)+\cos \left(\gamma _1-\gamma
   _2\right)-2 \cos \left(\gamma _2\right)-\sqrt{3} \sin \left(\gamma
   _1\right)-\sqrt{3} \sin \left(\gamma _1-\gamma _2\right)+6\right)}.
\end{align}
All probabilities are found. Moreover we find that $p^b_{\beta_l}=1/3,\; l=1,2,3$ by inserting \eqref{sju} in \eqref{elva} or directly by \eqref{CBORN}.
We let $\gamma_1=\gamma_2$ in order to get more compact expressions. This leads to
\begin{align}
p_{\beta_1\alpha_{12}}&=\frac{4 \sin \left(\gamma _2+\frac{\pi }{6}\right)-5}{3 \left(2 \sin \left(\gamma _2+\frac{\pi
   }{6}\right)-7\right)},\;
p_{\beta_1\alpha_{13}}=\frac{2}{3},\;
p_{\beta_1\alpha_{23}}=\frac{4 \sin \left(\gamma _2+\frac{\pi }{6}\right)-5}{3 \left(2 \sin \left(\gamma _2+\frac{\pi
   }{6}\right)-7\right)}\\\nonumber
p_{\beta_2\alpha_{12}}&=-\frac{-5 \cos \left(\gamma _2\right)+\sqrt{3} \sin \left(\gamma _2\right)+8}{3 \left(2
   \sin \left(\gamma _2+\frac{\pi }{6}\right)-7\right)},\;
p_{\beta_2\alpha_{13}}=\frac{1}{6},\;
p_{\beta_2\alpha_{23}}=\frac{2 \left(-2 \cos \left(\gamma _2\right)+\sqrt{3} \sin \left(\gamma
   _2\right)-4\right)}{3 \left(2 \sin \left(\gamma _2+\frac{\pi }{6}\right)-7\right)}\\\nonumber
p_{\beta_3\alpha_{12}}&=\frac{2 \left(-2 \cos \left(\gamma _2\right)+\sqrt{3} \sin
   \left(\gamma _2\right)-4\right)}{3 \left(2 \sin \left(\gamma _2+\frac{\pi }{6}\right)-7\right)},\;
p_{\beta_3\alpha_{13}}=\frac{1}{6},\;
p_{\beta_3\alpha_{23}}=-\frac{-5 \cos \left(\gamma
   _2\right)+\sqrt{3} \sin \left(\gamma _2\right)+8}{3 \left(2 \sin \left(\gamma _2+\frac{\pi
   }{6}\right)-7\right)}.
\end{align}
Here we calculate $\lambda_{i,jk}$ when $\gamma _2=\gamma _1$
$$ \lambda_{1,12}=\lambda_{1,23}=-\frac{\sqrt{1+\sin \left(\gamma _2+\frac{\pi }{6}\right)}}{ \sqrt{ 10-8 \sin \left(\gamma _2+\frac{\pi }{6}\right)}}, \; \lambda_{1,13}=1$$
$$\lambda_{2,12}=\lambda_{3,23}=\frac{-4 \cos \left(\gamma _2\right)+2 \sqrt{3} \sin \left(\gamma _2\right)+1}{2 \sqrt{4 \sin \left(\frac{\pi }{6}-2
   \gamma _2\right)+2 \sin \left(\gamma _2+\frac{\pi }{6}\right)+6}},\;\lambda_{2,13}=\lambda_{3,12}=-\frac{1}{2}$$
   $$\lambda_{2,23}=\lambda_{3,12}=-\frac{-5 \cos \left(\gamma _2\right)+\sqrt{3} \sin \left(\gamma _2\right)-1}{2 \sqrt{4 \sin \left(\frac{\pi }{6}-2
   \gamma _2\right)+2 \sin \left(\gamma _2+\frac{\pi }{6}\right)+6}}$$
We calculate numerically the extreme values for all $ \lambda $ to prove the existence of angles between $ 0 $ and $ 2 \pi $ for $ \arccos\lambda $. Thus, calculate $\frac{d}{d \gamma_2}\lambda=0$ and the limits when ${\gamma_2\rightarrow \pm4\pi/3}$, 
$$\frac{d}{d \gamma_2}\lambda_{1,12}=\frac{d}{d \gamma_2}\lambda_{1,23}=0\Rightarrow \gamma _2=\frac{\pi }{3}+2 \pi  C,\;C\in \mathbf{Z}$$ and $\lambda_{1,12}=\lambda_{1,23}=-1, \gamma _2=\frac{\pi }{3}+2 \pi  C$.
The problems arise when $\gamma_2\rightarrow 4\pi/3 $ for $\lambda_{1,12}$ and $\lambda_{1,23}$, since $ \lim_{\gamma_2\rightarrow \pm4\pi/3}=0$. Let $\lambda_{1,12}=\lambda_{1,23}=0$ when $\gamma_2= 4\pi/3$.
Moreover there exist angles
\begin{align}
\pi/2 \leq\varphi_{\beta_1,\alpha_1}-\varphi_{\beta_1,\alpha_2}\leq \pi,\\\nonumber
\pi/2 \leq\varphi_{\beta_1,\alpha_2}-\varphi_{\beta_1,\alpha_3}\leq \pi,
\end{align}
 where $\arccos{\lambda_{1,12}}=\varphi_{\beta_1,\alpha_1}-\varphi_{\beta_1,\alpha_2}$.
We see that the denominator of $ \lambda_{2,23},\;\lambda_{3,23},\;\lambda_{3,12},\;\lambda_{2,12}$ goes to zero when $ \gamma _2\to 4\pi /3\pm$. We therefore examine the limits
\begin{equation}\label{MinL}\lim_{\gamma_2\to 4\pi /3\pm} \, \lambda_{3,23}=\lim_{\gamma _2\to 4\pi /3\pm} \, \lambda_{2,12}=\lim_{\gamma _2\to 4\pi /3\mp} \, \lambda_{2,23}=\lim_{\gamma _2\to 4\pi /3\mp} \, \lambda_{3,12}=\pm\frac{\sqrt{3}}{2}.
\end{equation}

Analysis of extreme values provides us with the following 
\begin{equation}\label{MaxL1}\frac{d}{d \gamma_2}\lambda_{2,23}=\frac{d}{d \gamma_2}\lambda_{3,12}=0\Rightarrow \gamma _2=2 \pi  C,\;C\in \mathbf{Z},
\end{equation}
 and
\begin{equation}\label{MaxL2}\frac{d}{d \gamma_2}\lambda_{3,23}=\frac{d}{d \gamma_2}\lambda_{2,12}=0\Rightarrow \gamma _2=\frac{2 \pi }{3}+2 \pi C,\;C\in \mathbf{Z}.\end{equation}
The minimum values are given by \eqref{MinL}, $\min\lambda_{3,23}=\min\lambda_{3,23}=\min\lambda_{3,23}=\min\lambda_{3,23}=-\frac{\sqrt{3}}{2},\; \gamma_2\to 4\pi /3\text{ or } \gamma_2\to -4\pi /3 $. 
The maximum values are given by \eqref{MaxL1} and \eqref{MaxL2},
$\max\lambda_{2,23}=\max\lambda_{3,12}=1,\; \gamma _2=2 \pi  C ,\;C\in \mathbf{Z}$.
$\max\lambda_{3,23}=\max\lambda_{2,12}=1,\; \gamma _2=\frac{2 \pi }{3}+2 \pi C;C\in \mathbf{Z}$.
This prove that there exist angles such that 
\begin{align}
0 \leq\varphi_{\beta_2,\alpha_2}-\varphi_{\beta_2,\alpha_3}\leq \frac{5 \pi }{6},\\\nonumber
0 \leq\varphi_{\beta_3,\alpha_2}-\varphi_{\beta_3,\alpha_3}\leq \frac{5 \pi }{6},\\\nonumber
0 \leq\varphi_{\beta_2,\alpha_1}-\varphi_{\beta_2,\alpha_2}\leq \frac{5 \pi }{6},\\\nonumber
0 \leq\varphi_{\beta_3,\alpha_1}-\varphi_{\beta_3,\alpha_2}\leq \frac{5 \pi }{6}.
\end{align}
and
\begin{align}
\varphi_{\beta_1,\alpha_1}-\varphi_{\beta_1,\alpha_3}=0,\\\nonumber
\varphi_{\beta_2,\alpha_1}-\varphi_{\beta_2,\alpha_3}=\frac{2 \pi }{3},\\\nonumber
\varphi_{\beta_3,\alpha_1}-\varphi_{\beta_3,\alpha_2}=\frac{2 \pi }{3}
\end{align}
\section{Triple slit experiment}
An interesting example of interplay of two incompatible trichotomous observables is given by 
the triple slit experiment -- a natural generalization of the well know two slit experiment.
There are given: a) a source of quantum systems which has  very low intensity (so it might be interpreted as single-particle
source); 2) a screen with three slits $\alpha=1,2,3$ and each of them  can be open or close
on the demand; c) registration screen; typically it is covered by photo-emulsion; this produces the continuous 
interefrence picture; we shall consider discrete experiment. The $a$-observable gives slit's which is so to say is 
passed by a particle on the way from the source to the registration screen. To measure $a,$ an experimenter puts three detectors directly behind  slits.  We set $a=\alpha_i,$  if the detector behind the $i$th slit produces a click. By the assumption the source has so low intensity that one can neglect by double clicks (e.g., the detectors never click  simultaneously). We can find probabilities $p_\alpha^a, \alpha=1,2,3.$ 
This is the first experiments producing $a$-probabilities. Now we consider basic experiments.

To make the second variable discrete, we put detectors in three fixed places of the registration slit. It gives us the observable $b=\beta, \beta =1,2,3.$ Thus $\beta=1$ if the first detector clicks and so on.
The experiment is repeated at a few incompatible contexts, see \cite{K8} for general presentation

\medskip

$C_{123}:$ all slits are open; probabilities of $b$ detection collected in this context are 
probabilities $p_\beta^b= P_\psi(b=\beta)$ from section \ref{HH7}. In the QM-formalism context $C_{123}$ 
is represented by a quantum state $\psi.$ 

$C_{\alpha}, \alpha=1,2,3:$ only the slit $\alpha$ is open; corresponding probabilities are 
$p_{\beta \alpha}^{b\vert a},$ transition probabilities. In the QM-formalism context $C_{\alpha}$ 
is represented by  the quantum state $e_\alpha^a.$  

$C_{\alpha_{ij}}, i\not= j:$ only slits $\alpha_i$ and $\alpha_j$ are open (so the slit $\alpha_k, k\not=i,j,$ 
is closed); corresponding probabilities are $\paaj p^{b|a}_{\beta_l\alpha_{ij}}.$ In the QM-formalism context $C_{\alpha_{ij}}, i\not= j$ 
is represented by the quantum state 
$\psi_{\alpha_{ij}}= \frac{(\hpaai+\hpaaj)\psi}{\left\vert\right\vert
(\hpaai+\hpaaj)\psi\left\vert\right\vert}.$  

\medskip

Thus all probabilities discussed in section \ref{HH7} can be collected in this experiment. 
It is possible to check whether these experimental probabilities match the predictions of 
QM. The easiest way for experimenters is to check Sorkin equality (\ref{CBORN}).

Recently the group of  Gregor Weihs performed the triple slit experiment\footnote{It is surprising that 
it has not been done for long ago!}, see \cite{Urbasi}. They cliam that Sorkin's equality and  Born's rule are violated
by their experimental statistical data.

\section{Construction of a complex probability amplitude satisfying Born's rule}\label{cbr}
Now we have a pair of trichotomous observables $a$ and $b$ 
taking values $a=\alpha_i,\;i=1,2,3$ and $b=\beta_l, l=1,2,3.$ We do not assume that they have any 
relation to quantum physics; e.g., these are some random variables observed in biology or finances. It is assumed that there are given probability distributions of these variables
\[p_{\beta_l}^b =P(b=\beta_l), \; p_{\alpha_i}^a= P(a= \alpha_i).\]
Thus
\begin{equation}
\label{HH}
\sum_{l=1}^3 p_{\beta_l}^b=1, \; 
\sum_{i=1}^3 p_{\alpha_i}^a=1.
\end{equation}  
It is also assumed that there are given  conditional probabilities 
$
p_{\beta_l \alpha_j}^{b \vert a} = P(b=\beta_l|a= \alpha_i).
$
We know that for any sort of data the matrix of transition probabilities is stochastic, i.e., for each 
$\alpha_i$
\begin{equation}
\label{HH1}
\sum_{l=1}^3  p_{\beta_l \alpha_i}^{b \vert a}= 1.
\end{equation}
Finally, we assume a possibility to collect the data on measurements of 
observables $\Paai, i=1, 2,3,$  probabilities
$
p_{\beta_l\alpha_{kj}}^{b|a} \equiv  P(b=\beta_l|\Paai=0).
$
``The detector corresponding to the value $a=\alpha_i$ does not click, so the value of 
$a$ is either $a=\alpha_j$ or $a=\alpha_k,$ where $j,k\not= i.$ However, we do not know 
the value of $a.$ In this context we measure the $b$-variable.''
Fo any sort of data, we have 
\begin{equation}
\label{HH2}
\sum_{l=1}^3 p_{\beta_l\alpha_{kj}}^{b|a}=1.
\end{equation}

\subsection{Complex amplitude matching Born's rule for one observable}

Now we want to find a complex probability amplitude $\psi_{\beta_l}$ such that 
Born's rule (for the $b$-variable) holds: $\vert \psi_{\beta_l} \vert^2= p_{\beta_l}^b.$
We  copy the QM-scheme, so we represent 
$
\psi_{\beta_l} = \psi_{\beta_l \alpha_1} + \psi_{\beta_l \alpha_2} + \psi_{\beta_l \alpha_3},
$
where the sub-amplitudes 
$
\psi_{\beta_l \alpha_i} = \sqrt{p^{a}_{\alpha_i}p^{b|a}_{\beta_l,\alpha_i}}e^{i \varphi_{\beta_l\alpha_i}}
$
and phases are determined by the system of equations (\ref{sju}). It is convenient to work with the 
{\it interference coefficients}, see \cite{K8},  given by right-hand sides of these equations
\begin{equation}
\label{sju1}
\lambda_{l, ij}= \lambda_{l, ji}=\frac{(\paai+\paaj)p_{\beta_l\alpha_{ij}}^{b|a}-(\paai p_{\beta_l\alpha_{i}}^{b|a}+\paaj p_{\beta_l\alpha_{j}}^{b|a})}{2\sqrt{\paai p_{\beta_l\alpha_{i}}^{b|a}\paaj p_{\beta_l\alpha_{j}}^{b|a}}}.
\end{equation}
Interference coefficients obtained in quantum physics are always bounded by 1:
\begin{equation}
\label{sju1j}
\vert \lambda_{l, ij} \vert \leq 1. 
\end{equation}
However, since we start with data of any origin, the condition (\ref{sju1j}) has to be checked
to proceed to representation of data by complex amplitudes.\footnote{If this condition is violated 
then data may be represented by so called hyperbolic probabilistic amplitudes \cite{P2}.} 
If the system of equations, $m=1,2,3,$
\begin{equation}
\label{sju2J}
\cos(\varphi_{\beta_m\alpha_1}-\varphi_{\beta_m\alpha_2})= \lambda_{m, 12}, 
\end{equation}
$$
\cos(\varphi_{\beta_m\alpha_2}-\varphi_{\beta_m\alpha_3}) = \lambda_{m, 23}, 
$$
$$
\cos(\varphi_{\beta_m\alpha_1}-\varphi_{\beta_m\alpha_3}) = \lambda_{m, 13}.
$$
has a solution (three phases) then we can construct the probability amplitudes $\psi_{\beta_l \alpha_i}$ and, hence, the probability
amplitudes $\psi_{\beta_l}$ and the corresponding vector $\psi.$ 

However, in general
such amplitudes will not provide a solution of the ``inverse Born problem'', namely, Born's rule can be violated.
To obtain the real solution one should solve the system (\ref{sju2J}) under the constriant (\ref{CBORN}).
Thus, to proceed toward a proper complex amplitude, one should first check the validity of (\ref{CBORN})
and then to solve the system (\ref{sju2J}). It is convenient to express ``triple probabilities'' $p_{\beta_m, \alpha_{ij}}$
through coefficients of interference 
\begin{equation}
\label{sju2k}
p_{\beta_m \alpha_{ij}}=\frac{1}{p_{\alpha_i} + p_{\alpha_j}}(p_{\alpha_i}p_{\beta_m\alpha_i} + p_{\alpha_j} p_{\beta_m\alpha_j} -
2 \lambda_{\beta_m \alpha_{ij}} \sqrt{p_{\alpha_i}p_{\beta_m\alpha_i}p_{\alpha_j} p_{\beta_m\alpha_j}}).
\end{equation}
We remark that if (\ref{sju1j}) holds, then triple probabilities given by   (\ref{sju2k}) are always nonnegative.
By using the $\lambda$-variables normalization equations (\ref{HH2}) can be written as $(j,k=1,2,3)$
\begin{equation}
\label{jjj}
\sum_{l=1}^3 \lambda_{l,jk} \sqrt{p_{\beta_l \alpha_j} p_{\beta_l \alpha_k}}=0.
\end{equation}
 We also can write Sorkin's equality (in fact, the formula of total probability 
with interference terms) as 
\begin{equation}
\label{elvah} p_{\beta_l}^b
=p_{\alpha_i}^a p_{\beta_l \alpha_i}^{b|a}+p_{\alpha_j}^a p_{\beta_l \alpha_j}^{b|a}+p_{\alpha_k}^a p_{\beta_l \alpha_k}^{b|a}
+2\lambda_{l,ij}\sqrt{p_{\alpha_i}p_{\alpha_j}p_{\beta_l \alpha_i}^{b|a}p_{\beta_l \alpha_j}^{b|a}}
\end{equation}
$$
+2\lambda_{l,ik}\sqrt{p_{\alpha_i}p_{\alpha_k}p_{\beta_l \alpha_i}^{b|a}p_{\beta_l \alpha_k}^{b|a}}
+2\lambda_{l,jk}\sqrt{p_{\alpha_j}p_{\alpha_k}p_{\beta_l \alpha_j}^{b|a}p_{\beta_l \alpha_k}^{b|a}}.
$$
Hence, to obtain Born's rule for the $b$-variable which matches the intereference formula of total probability, we have to  find $\lambda$ satisfying equations 
(\ref{jjj}) and (\ref{elvah}) and put such $\lambda$ into equations  (\ref{sju2J}), then solve this system of equations.
In general, it is a complex problem.

Thus, finally, we can write the complete system of equations:
\begin{equation}
\label{jjj1}
\sum_{l=1}^3 \lambda_{l,jk} \sqrt{p_{\beta_l \alpha_j} p_{\beta_l \alpha_k}}=0, \; j,k=1,2,3;
\end{equation}

\begin{equation}
\label{elvah1} p_{\beta_l}^b
=p_{\alpha_i}^a p_{\beta_l \alpha_i}^{b|a}+p_{\alpha_j}^a p_{\beta_l \alpha_j}^{b|a}+p_{\alpha_k}^a p_{\beta_l \alpha_k}^{b|a}
+2\lambda_{l,ij}\sqrt{p_{\alpha_i}p_{\alpha_j}p_{\beta_l \alpha_i}^{b|a}p_{\beta_l \alpha_j}^{b|a}};
\end{equation}
$$
+2\lambda_{l,ik}\sqrt{p_{\alpha_i}p_{\alpha_k}p_{\beta_l \alpha_i}^{b|a}p_{\beta_l \alpha_k}^{b|a}}
+2\lambda_{l,jk}\sqrt{p_{\alpha_j}p_{\alpha_k}p_{\beta_l \alpha_j}^{b|a}p_{\beta_l \alpha_k}^{b|a}}.
$$

\begin{equation}
\label{sju2J1}
\cos(\varphi_{\beta_m\alpha_1}-\varphi_{\beta_m\alpha_2})= \lambda_{m, 12}, 
\end{equation}
\begin{equation}
\label{sju2J2}
\cos(\varphi_{\beta_m\alpha_2}-\varphi_{\beta_m\alpha_3}) = \lambda_{m, 23}, 
\end{equation}
\begin{equation}
\label{sju2J3}
\cos(\varphi_{\beta_m\alpha_1}-\varphi_{\beta_m\alpha_3}) = \lambda_{m, 13}.
\end{equation}

Solution of this system will provide us a complex probability amplitude $\psi$
such that $\vert \langle \psi \vert e_{\beta}^b\rangle \vert^2 = p_\beta^b.$

Let us consider the case of maximally unbiased matrix of transition probabilities;
\begin{equation}\label{i1}
p^{a|b}_{\beta_l\alpha_j}=1/3,\; \forall\; l,j=1,2,3
\end{equation}
Moreover, to simplify the task by the factor of three, we will put all 
\begin{equation}\label{i2}
p^{b}_{\beta_l}=1/3,\; \forall\; l=1,2,3
\end{equation}
Now, let us introduce new variables $x>0$ and $y>0$:
\begin{equation}\label{i3}
\sqrt{p^{a}_{\alpha_l}/p^{a}_{\alpha_2}}=x,\; \sqrt{p^{a}_{\alpha_l}/p^{a}_{\alpha_3}}=y,\;
\sqrt{p^{a}_{\alpha_2}/p^{a}_{\alpha_3}}=y/x,
\end{equation}
That means that 
\begin{equation}\label{i4}
p^{a}_{\alpha_l}=\frac{x^2 y^2}{y^2 x^2+x^2+y^2},\;
p^{a}_{\alpha_2}=\frac{ y^2}{y^2 x^2+x^2+y^2},\;
p^{a}_{\alpha_3}=\frac{x^2 }{y^2 x^2+x^2+y^2},
\end{equation}
and the condition $p^{a}_{\alpha_l}+p^{a}_{\alpha_2}+p^{a}_{\alpha_3}=1$
always holds.
Let proceed for a particular choice of interference coefficients (ansatz) $\lambda_{l,12}=\mu,\;\lambda_{l,13}=-\mu,$, thus $\lambda_{l,23}=1-2\mu^2$  by \begin{equation}\label{i25}
\lambda_{l,23}=\lambda_{l,12}\lambda_{l,13}\pm\sqrt{(1-\lambda_{l,12}^2)(1-\lambda_{l,13}^2)}.
\end{equation}
The system of equations \eqref{sju1} for $\lambda$  under conditions \eqref{i1} and \eqref{i2} have the form:
\begin{align}\label{i7}
\lambda_{l,12}=\frac{3}{2}\left(x+\frac{1}{x}\right)\left(p_{\beta _l\alpha _{12} }^{b|a}-\frac{1}{3}\right)\\\nonumber
\lambda_{l,13}=\frac{3}{2}\left(y+\frac{1}{y}\right)\left(p_{\beta _l\alpha _{13} }^{b|a}-\frac{1}{3}\right)\\\nonumber
\lambda_{l,23}=\frac{3}{2}\left(\frac{y}{x}+\frac{x}{y}\right)\left(p_{\beta _l\alpha _{23} }^{b|a}-\frac{1}{3}\right).
\end{align}
We write this as:
\begin{equation}\label{i6}
p^{b|a}_{\beta_l\alpha_{12}}=\frac{1}{3}+\frac{2 \mu }{3 \left(x+\frac{1}{x}\right)},\;
p^{b|a}_{\beta_l\alpha_{13}}=\frac{1}{3}-\frac{2 \mu }{3 \left(y+\frac{1}{y}\right)},\;
p^{b|a}_{\beta_l\alpha_{23}}=\frac{1}{3}+\frac{2 \left(1-2 \mu ^2\right)}{3
   \left(\frac{x}{y}+\frac{y}{x}\right)},
\end{equation}
where $\mu$ is a parameter.
\begin{figure}[ht] \centering
\includegraphics[width=84.582mm]{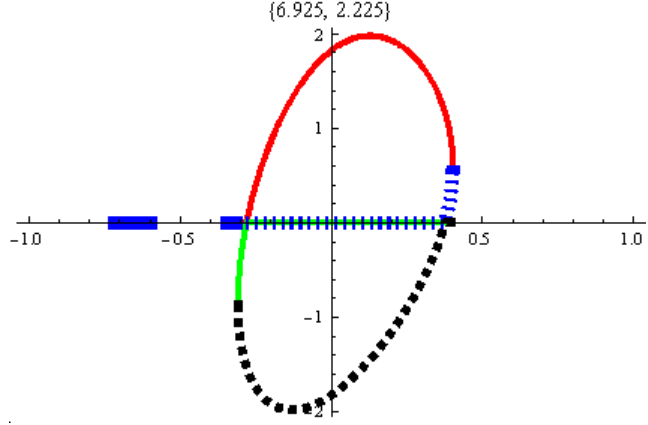}
\caption{Left side of the expression (\ref{i12}) is plotted versus parameter $v=\lambda_{l,13}$ for the fixed values of $x=6.925, y=2.225$ and different signs before square roots in expressions for $\lambda_{l,23},$ see (\ref{i23}), and $\lambda_{l,12},$
see  (\ref{i13}). 
We can see, that we can often satisfy (\ref{i12}) by properly choosing signs in expressions (\ref{i23})  and (\ref{i13}).}
\end{figure}
The probabilities given by \eqref{i6} satisfy the relation in \eqref{CBORN}
which in this case looks as:
\begin{equation}\label{i8}(x^2+1)y^2 p_{\beta _l\alpha _{12} }+(y^2+1)x^2 p_{\beta _l\alpha _{13} }+(x^2+y^2)p_{\beta _l\alpha _{23} }=\frac{2}{3}(x^2y^2+y^2+x^2)
\end{equation}
Putting \eqref{i6} into \eqref{i8}, we get an equation for $\mu$:
\begin{equation}\label{i9}
2\mu^2+(x-y)\mu-1=0\Rightarrow \mu=\frac{(y-x)\pm\sqrt{(x-y)^2+8}}{4}
\end{equation}
We are interested in the case then all absolute value of lambdas are less than one 
\begin{equation}\label{i11}|\lambda_{l,{ij}}|<1, \forall \;ij=12,13,23.\end{equation}
It is satisfied when $|\mu|<1$. So, for the case of $|x-y|<1$
, for  both roots of  \eqref{i9} conditions $|\lambda_{l,{ij}}|<1$ are valid, 
if $x-y>1$, then \eqref{i9} with the plus sign suits $|\lambda_{l,{ij}}|<1$, 
otherwise $y-x>1$, then \eqref{i9} with the minus sign is valid.
\\ 
Now we proceed in a general case, i.e. without ansatz $\lambda_{l,12}=\mu,\;\lambda_{l,13}=-\mu,\;\lambda_{l,23}=1-2\mu^2$.  Conditions \eqref{i1} and \eqref{i2} equation \eqref{CBORN}, which is equivalent to Born's rule, comes down to:
\begin{equation}\label{i12}
y\lambda_{l,12}+x\lambda_{l,13}+\lambda_{l,23}=0
\end{equation}
We should combine it with the constraint, see \eqref{i25} for $\lambda_{l,12},\lambda_{l,13},\lambda_{l,23}$
 to have simultaneous solution \begin{equation}\label{i23}
\lambda_{l,23}=\lambda_{l,12}\lambda_{l,13}\pm\sqrt{(1-\lambda_{l,12}^2)(1-\lambda_{l,13}^2)}
\end{equation}
We have two equations for three variables, thus we can express the solution as a one-parametric family. Let us choose $v=\lambda_{l,13}$ as a parameter.
Then 
\begin{equation}\label{i13}
\lambda_{l,12}=\frac{-xv(y+v)\pm\sqrt{(v^2-1)(v^2x^2-y^2-1-2yv)}}{y^2+1+2yv},
\end{equation}and
$\lambda_{l,23}$
 can be obtained from equation \eqref{i25}.
We have to make sure that $\lambda_{l,12},\lambda_{l,13},\lambda_{l,23}$
 exist, are real and satisfy \eqref{i11}, given real and positive \textit{x} and \textit{y}.
 In this, more general, case
\begin{equation}\label{i14}
p^{b|a}_{\beta_l\alpha_1}=\frac{1}{3}+\frac{2 \lambda_{l,12} }{3 \left(x+\frac{1}{x}\right)},\;
p^{b|a}_{\beta_l\alpha_1}=\frac{1}{3}+\frac{2 \lambda_{l,13} }{3 \left(y+\frac{1}{y}\right)},\;
p^{b|a}_{\beta_l\alpha_1}=\frac{1}{3}+\frac{2  \lambda_{l,23}}{3
   \left(\frac{x}{y}+\frac{y}{x}\right)},
\end{equation}
Seeing that all values in the parentheses in \eqref{i6} are greater than 2, each of the this probabilities non-negative and smaller than 2/3, if $\lambda_{l,12},\lambda_{l,13},\lambda_{l,23}$ are smaller than 1.
The main problem is to describe possible rangers of parameters in \eqref{i13} which give us $|\lambda_{l,ij}|\leq 1$, see figure 1--3. W e remark that $\lambda_{l,13}$ is a a parameter, $|v|\leq 1$. 

\begin{figure}[ht]\centering
\includegraphics[width=85.725mm]{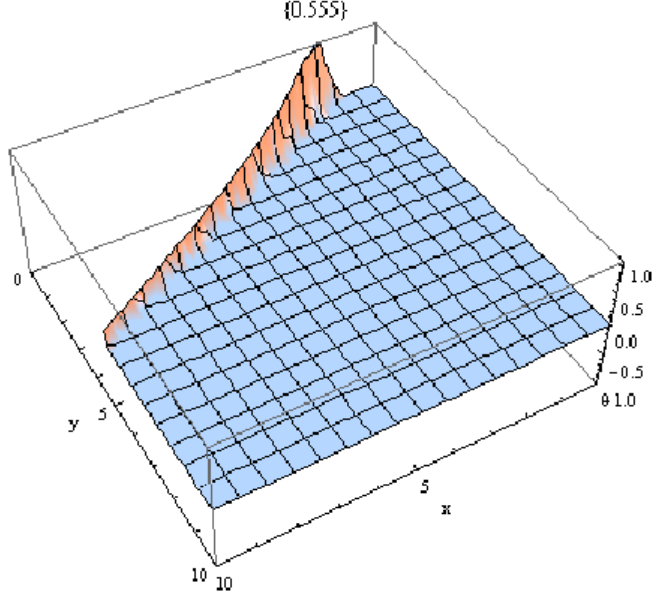}
\caption{Left side of the expression (\ref{i12}) is plotted versus parameters $x, y,$ for the fixed $v=\lambda_{l,13}=0,555.$ Minus sign is taken before square root in (\ref{i13}), plus sign is taken in (\ref{i23}) when obtaining $\lambda_{l,13}$ values.
We can see that it is equal to zero as (\ref{i12}) demands on a large scope of $x, y$ parameters values.}
\end{figure}
\begin{figure}[ht]\centering\large 
\includegraphics[width=97.282mm]{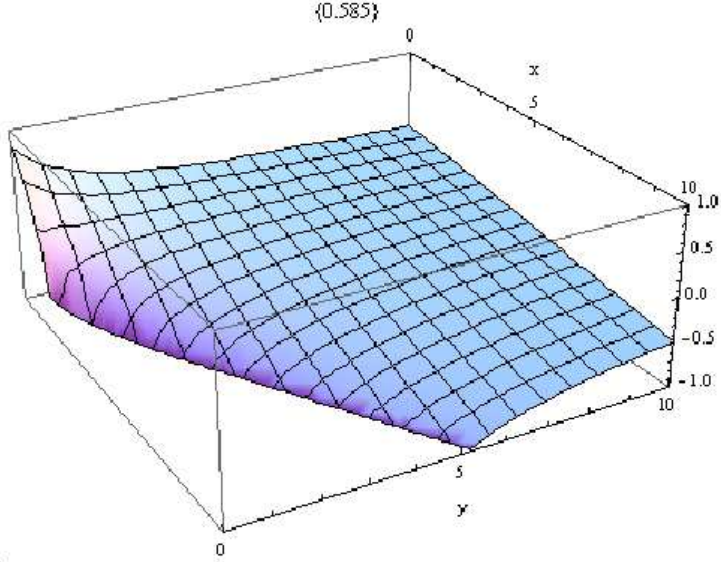}
\caption{Value of $\lambda_{l,12}$ as function (\ref{i13}) of parameters $x, y,$ for the fixed $v=\lambda_{l,13}=0.585.$ Plus sign is taken before square root in \ref{i13}).
One may see that value $\lambda_{l,12}$ is within $(-1,1)$ limits for the most part of the values $x$ and $y.$ }
\end{figure}

\subsection{Complex amplitude matching Born's rule for two observables}

We now want to select an orthonormal basis
$e_{\alpha}^a$ such that, for the state $\psi$ constructed in the previous section,
$\vert \langle \psi \vert e_{\alpha}^a\rangle \vert^2 = p_\alpha^a.$
We turn to considerations of section \ref{Un}. 
Since vectors of this basis can be selected up to $e^{i \xi_\alpha}.$ We can
select $\theta_{\beta_1 \alpha_k}= \phi_{\beta_1 \alpha_k},$ i.e., 
set $\xi_\alpha=0.$ 
Of course, to guarantee orthogonality of this basis, constraints (\ref{UI1H}), (\ref{UI2H})
should be taken into account: 

\begin{equation}
\label{UI1Hh}
 \sum_m \sqrt{p_{\beta_m \alpha_i}p_{\beta_m \alpha_j}} \lambda_{m, ij}=0,
 \end{equation}
\begin{equation}
\label{UI2Hh}
 \sum_m \epsilon_{m, ij} \sqrt{p_{\beta_m \alpha_i}p_{\beta_m \alpha_j}}  \sqrt{1- \lambda_{m, ij}^2}= 0,
 \end{equation}
where $\epsilon_{m, ij}$ are signs which are selected in a proper way.

Moreover, the matrix of transition probabilities has to be doubly stochastic, i.e.,  
besides (\ref{HH1}), we should have
\begin{equation}
\label{HH3}
\sum_{l=j}^3  p_{\beta_l \alpha_j}^{b \vert a}=1
\end{equation}
for each $l=1,2,3.$

Under these conditions the complex amplitude $\psi$ produced by our algorithm 
matches with Born's rule for both obsrevables, $a$ and $b.$ \\


{\bf Example 1.} We can take 
\[\lambda_{1, 12}= \mu, \; \lambda_{1, 23}= - \mu, \lambda_{1, 13}= 0;\]
\[\lambda_{2, 12}= - \mu, \lambda_{2, 23}= 0, \; \lambda_{2, 13}= \mu;\]
\[\lambda_{3, 12}= 0, \; \lambda_{3, 23}= \mu, \; \lambda_{3, 13}=  - \mu.\]

This gives us the solution $\mu= \frac{1}{\sqrt{2}}.$ First take the plus-case(e.i. $\mu=\pm \frac{1}{\sqrt{2}}$). 
We select $\phi_{m1}=\nu_m, m=1,2,3.$
For $\beta_1,$ we obtain the system of equations: $\cos(\nu_1-\phi_{12})=\frac{1}{\sqrt{2}},
\cos(\phi_{12} - \phi_{13})=- \frac{1}{\sqrt{2}}, \cos(\nu_1-\phi_{13})=0.$
Hence,  $\phi_{11}=\nu_1,$ and  
$\phi_{12}=\nu_1- \pi/4,  \phi_{13}=\nu_1+\pi/2$ or $\phi_{12}=\nu_1+\pi/4,  \phi_{13}=\nu_1-\pi/2.$

Then, for $\beta_2,$ $\cos(\nu_2-\phi_{22})=- \frac{1}{\sqrt{2}},
\cos(\phi_{22} - \phi_{23})= 0, \cos(\nu_2-\phi_{23})= \frac{1}{\sqrt{2}}.$
Hence,  $ \phi_{21}=\nu_2,$ and  $\phi_{22}=\nu_2+3\pi/4,  \phi_{23}=\nu_2+ \pi/4$ or
$\phi_{22}=\nu_2-3\pi/4,  \phi_{23}= \nu_2- \pi/4.$
Finally, for  $\beta_3,$ $\cos(\nu_3-\phi_{32})=0,
\cos(\phi_{32} - \phi_{33})= \frac{1}{\sqrt{2}}, \cos(\nu_3-\phi_{33})= - \frac{1}{\sqrt{2}}.$
Hence,  $ \phi_{31}=\nu_3,$ and  $\phi_{32}=\nu_3+\pi/2,  \phi_{33}=\nu_3+ 3\pi/4$ or    
$\phi_{32}=\nu_3- \pi/2,  \phi_{33}=\nu_3- 3\pi/4.$
We have from equation \eqref{RRTT3},\eqref{RRTT4} and \eqref{tio} that
$$\psi_{\beta_1}=\sum_{i}\sqrt{p_{\alpha_m}^a p_{\beta_l \alpha_m}}e^{i( \xi_{\alpha_i}+\theta_{\beta_l\alpha_m})},$$ where $\xi_{\alpha_i}=0$ and $\sqrt{p_{\alpha_m}^a p_{\beta_l \alpha_m}}=\frac{1}{3}.$
Thus $$\psi_{\beta_1}= \frac{1}{3}(e^{i \nu_1}+e^{i \nu_1\mp \pi/4}+e^{i\nu_1 \pm\pi/2})=\frac{e^{i \nu_1}}{3}(1+e^{\mp i \pi/4}+e^{\pm i\pi/2})=\frac{e^{i \nu_1}}{3}((1+ \frac{\sqrt{2}}{2}) \pm i (1 - \frac{\sqrt{2}}{2}));$$
$\psi_{\beta_2}= \frac{e^{i \nu_2}}{3}(1 \pm i \sqrt{2})$ and, finally, 
$ \psi_{\beta_3}=\frac{e^{i \nu_3}}{3}((1- \frac{1}{\sqrt{2}}) \pm i (1 + \frac{1}{\sqrt{2}})).$ 
We remark that $\vert \psi_{\beta_j}\vert^2= 1/3, j=1,2,3.$

QLRA produces following possible realizations of the ``wave function'':
\begin{equation}
\label{J1}
\psi= \frac{1}{3}[ ((1+ \frac{1}{\sqrt{2}}) \pm i (1 - \frac{1}{\sqrt{2}})) e_{\beta_1}^{b*} +(1 \pm i \sqrt{2}) e_{\beta_2}^{b}
+((1- \frac{1}{\sqrt{2}}) \pm i (1 + \frac{1}{\sqrt{2}}))e_{\beta_3}^{b*}],
\end{equation}
where $e_{\beta}^{b}$ are the orthonormal basis
\begin{equation}
\label{L2b}
e^{b}_{\beta_1}=\left(
\begin{array}{ll}
 e^{i \nu_1}\\
 0\\
 0
 \end{array}
 \right ), 
 \; e^{b}_{\beta_2}=\left(
\begin{array}{ll}
 0\\
 e^{i \nu_2}\\
 0
 \end{array}
 \right ),\; 
e^{b}_{\beta_3}=\left(
\begin{array}{ll}
 0\\
 0\\
 e^{i \nu_3}
 \end{array}
 \right ).
\end{equation}
For the minus-case (e.i. $\mu= -\frac{1}{\sqrt{2}}$) QLRA produces following ``wave function'':
\begin{equation}
\label{J2}
\psi= \frac{1}{3}[ ((1- \frac{1}{\sqrt{2}}) \pm i (1 + \frac{1}{\sqrt{2}})) e_{\beta_1}^{b} +(1 \pm i \sqrt{2}) e_{\beta_2}^{b}
+((1+ \frac{1}{\sqrt{2}}) \pm i (1 - \frac{1}{\sqrt{2}}))e_{\beta_3}^{b}].
\end{equation}
\subsection*{Acknowledgments}
One of the authors, Irina Basieva, is supported by Swedish Institute, post--doc 
fellowship. We would like to thank professor Andrei Khrennikov for fruitful discussions about quantum formalism and probability theory.



\end{document}